\documentclass[prl,10pt,twocolumn]{revtex4-1}
\usepackage{amssymb,amsmath,graphicx,color}
\usepackage{times}
\usepackage{bm}

\begin{document}

%\linenumbers
\def\sgn{\mathop{\rm sgn}}
	
\title{Synthetic spin-orbit coupling and topological polaritons in Janeys-Cummings lattices}

\author{Feng-Lei Gu$^1$, Jia Liu$^1$, Feng Mei$^{2,3}$, Suotang Jia$^{2,3}$, Dan-Wei Zhang$^1$ and Zheng-Yuan Xue$^1$}

\affiliation{$^1$Guangdong Provincial Key Laboratory of Quantum Engineering and Quantum Materials, and School of Physics\\ and Telecommunication Engineering, South China Normal University, Guangzhou 510006, China;\\ $^2$State Key Laboratory of Quantum Optics and Quantum Optics Devices, 		Institute of Laser Spectroscopy, Shanxi University, Taiyuan, Shanxi 030006, China and\\ $^3$Collaborative Innovation Center of Extreme Optics,		Shanxi University, Taiyuan, Shanxi 030006, China\\
Correspondence: Feng Mei (meifeng@sxu.edu.cn) or  Zheng-Yuan Xue (zyxue83@163.com).}

%\date{\today}

	\begin{abstract}
The interaction between a photon and a qubit in the Janeys-Cummings (JC) model generates a kind of quasiparticle called polariton. While they are widely used in quantum optics, difficulties in  engineering controllable coupling of them severely limit their applications to simulate spinful quantum systems. Here we show that, in the superconducting quantum circuit context, polariton states in the single-excitation manifold of a JC lattice can be used to simulate a spin-1/2 system, based on which tunable synthetic spin-orbit coupling and novel topological polaritons can be generated and explored. The lattice is formed by a sequence of coupled transmission line resonators, each of which is connected to a transmon qubit. Synthetic spin-orbit coupling and effective Zeeman field of the polariton can both be tuned by modulating the coupling strength between neighbouring resonators, allowing for the realization of a large variety of polaritonic topological semimetal bands. Methods for detecting the polaritonic topological edge states and topological invariants are also proposed. Therefore, our work suggests that the JC lattice is a versatile platform for exploring spinful topological states of matter, which may inspire developments of topologically protected quantum optical and information processing devices.
	\end{abstract}


	\maketitle

\section{Introduction}

The Janeys-Cummings (JC) model proposed in 1963 \cite{JCModel} is a seminal theoretical model treating light-matter interaction with full quantum theory, i.e., the interaction of a quantized electromagnetic field with a two-level atom. This model has been widely applied to many quantum platforms for studying the interaction of a quantized bosonic field with a qubit, which now has become the cornerstone in quantum optics and quantum computation \cite{Ion,CQED,SC3,SC5,Nano,Opto}. Furthermore, an interconnected array of multiple JC systems can form a JC lattice \cite{Hartmann,Fazio,Nori,Koch}, which provides an innovative quantum optical platform for studying condensed matter physics. This is highlighted by previous works which show that coupled JC systems can be used to realize the Bose-Hubbard model and investigate superfluid-to-Mott-insulator phase transition \cite{Hartmann2006,Greentree,Angelakis}.  However, spinful lattices have not been simulated in this platform due to the difficulty in engineering a tunable coupling between different cavities.
	
	On the other hand, the search of topological states of matter in artificial systems recently has become a rapidly growing field of research \cite{add-njp,add-prl109,TPCA1,TPCA2,TPCA3,TPPho1,TPPho2,TPPhonon,TPPolariton,TPOpto}. Topological states are characterized by topological invariants which are robust to the smooth changes in system parameters and disorders, where topological edge states can be employed for robust quantum transport \cite{Kane,Zhang}. Therefore, they hold tremendous promise for fundamental new states of matter as well as for dissipationless quantum transport devices and topological quantum computation \cite{TPQC}. One of the key ingredients for generating such states is to realize tunable spin-orbit coupling (SOC). Significant theoretical and experimental progress on realizing synthetic SOC recently have been achieved in ultracold atom systems \cite{SOC1,SOC2,SOC3}. This progress stimulate great research interests to explore topological states with ultracold atoms trapped in optical lattices \cite{TPExp2,TPExp3,TPExp4,TPExp6}. However, the limited trapping time and the site addressing difficulty increase the experimental complexity, i.e., it is generally difficult to have modulable coupling between two neighbouring sites in an optical lattice.

Here, we find that the JC lattice system can be used to realize various topological spin lattice models, where synthetic polaritonic SOC and Zeeman field can be induced with \emph{in situ} tunability, which provides a flexible platform to explore topological states of matter with great controllability. Specifically, we consider realizing the JC lattice in the context of superconducting quantum circuits, where each JC lattice site is constructed by a  transmission line resonator (TLR) coupled to a two-level transmon qubit. We find that the dressed polariton states in the single-excitation manifold in each JC lattice site can simulate a spin-1/2 system. Particularly, synthetic SOC and Zeeman field for polaritons can be induced and manipulated by only engineering the coupling strength between neighbouring resonators. Meanwhile, we show that, based on tunable synthetic SOC and Zeeman field, nodal-loop semimetal bands \cite{NLS, NodalLoop1, NodalLoop2} and topological polaritons can be realized and explored in the simulated JC lattice. Moreover, through calculating the topological winding number, we find that this tunable system has a rich topological phase diagram.
	
	Our proposal to explore the topological states in the JC lattice system is different from previous ones based on the optical lattices \cite{SOC1,SOC2,SOC3}. In particular, our proposal has a number of advantages. (i) Unlike ultracold atoms, polaritons are quasiparticles which are hybrids of photons and qubit excitations. Topological polaritons emerge from the topological structure of light-matter interaction, where photons and qubit excitations are topologically trivial by themselves, but combining together, they become hybrid topological states. Therefore using polariton for quantum simulation enriches our controlling methods -- both photonic and atomic means take effects. (ii) The systematic parameters in JC lattice systems can be tuned at a single-site level, which allows us to generate a wide variety of SOC forms. (iii) The geometry of JC lattice can be artificially designed and lattice boundaries are easy to be created for observing topological edge states, thus various topological lattice models and topological effects can be constructed and probed. (iv) The particle number putting in a JC lattice can be deterministically controlled. With such an advantage, we present a method using single-particle quantum dynamics to probe topological winding numbers and topological polariton edge states. (v) In the quantum optics platform, JC lattice systems previously have generated multiple important applications, including masers, lasers, photon transistors, and quantum information processors. Meanwhile, there are indeed several disadvantages in our proposed JC lattice system, such as the limited system size, parameter fluctuations, and decoherence. However, the essential physics of the simulated topological polariton states, such as the topological invariants and edge states, can still be detected under these realistic circumstances. Therefore, the topological JC lattice system in superconducting quantum circuits offers the possibility to develop functional topological spin quantum devices.

\begin{figure}[tb]
\centering
\includegraphics[width=6cm]{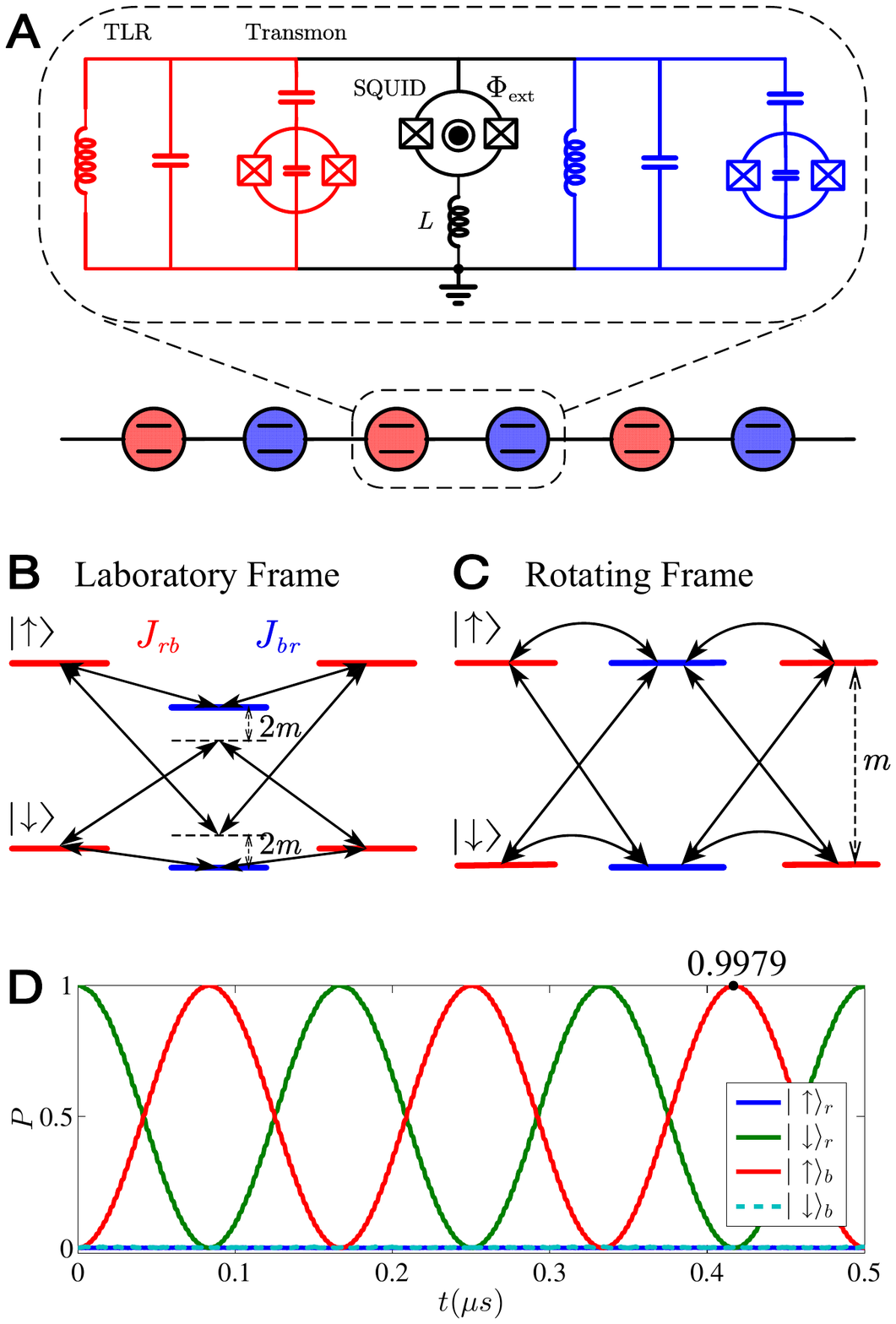}
\caption{\textbf{The proposed superconducting circuit implementation of spin-1/2 lattice models.} (A) The ``spin-1/2" polariton lattice with two types of unit cells, A-type (red) and B-type (blue), arranged alternately. Each unit cell has two pseudo-spin-1/2 states simulated by the two single-excitation eigenstates of the JC model. The two types of unit cells are of the different qubit and photon eigenfrequencies and JC coupling strengths.  The zoom-in figure details the equivalent superconducting circuits of two neighbouring unit cells and their coupling circuit, which is a combination of a SQUID and an inductor $L$ in series, to induce the tunable inter-cell photon hopping.   (B) The resonant and detuning couplings of inter-cell spin states. Since the alternate A- and B- type unit cells arrangement, two sets of driving, $J_{\text{AB}}$ and $J_{\text{BA}}$, have to be adopted to ensure the translation symmetry in the rotating frame defined by $U$. (C) The levels and designed hopping of the polariton lattice in the rotating frame, where the A-type and B-type unit cells can be treated as the same, so that the proposed circuit simulates a 1D spin-1/2 tight-binding lattice model. (D) The Rabi oscillation of two-unit-cell system to justify the treatment of the proposed inter-cell coupling. The considered transition $|\uparrow \rangle_{\text{A}}\leftrightarrow |\downarrow \rangle_{\text{B}}$ is of the worst meeting the RWA requirement among the 4 possible transitions, thus the fidelity obtained is the least one, but it still reaches a very high value of $0.9979$ in the third Rabi cycle. All the numerical simulations are based on the Hamiltonian in Eq. (\ref{H_JC}) without RWA.}
		\label{setup}
\end{figure}

\section{Results}
\subsection{Janeys-Cummings lattice}
	
The method for implementing a one-dimensional (1D) JC lattice in superconducting quantum circuits \cite{SC3} is as follow. As shown in Fig.~1A, every unit cell consists of a TLR resonantly coupled with a transmon, forming a JC model \cite{Nori-rew-Simu2-JC}. The neighbouring TLRs are connected by a combination of a SQUID and a small inductor $L$ in series, which can actually be regarded as the counterpart of a semitransparent mirror in the cavity QED system, allowing photons to hop across neighbouring cavities (see Methods). As a result, setting $\hbar= 1$ hereafter, the system Hamiltonian of this JC lattice is
	\begin{equation} \label{H_JC}
	H_{\text{JC}}=\sum_{l=1}^{N}h_l+\sum_{l=1}^{N-1}J_{l}(t)\left(\hat{a}_l^\dagger \hat{a}_{l+1}+\text{h.c.}\right)+H_{\text{c}},
	\end{equation}
where $N$ is the number of the unit cells; $h_l=  \omega_l (\sigma_l^+\sigma_l^- + \hat{a}_l^\dagger \hat{a}_l) + g_l\left(\sigma_l^+ \hat{a}_l + \sigma_l^- \hat{a}_l^\dagger\right)$ is the JC type interacting Hamiltonian in $l$th unit cell. 	The condition $g_l \ll  \omega_l$ has to be met for justifying the JC coupling. $\sigma_l^{+}=|\text{e}\rangle\langle \text{g}|$ and $\sigma_l^{-}=|\text{g} \rangle\langle \text{e}|$ are the raising and lowering operators of the $l$th transmon qubits. $\hat{a}_l$ and $\hat{a}_l^\dagger$ are the annihilation and creation operators of the photon in $l$th TLR. And $J_{l}(t)$ is the inter-TLR hopping strengths between $l$th and $(l+1)$th unit cells. Different from optical cavities, the time-dependence of $J_{l}(t)$ here can be induced by adding a time-varying external magnetic flux threading through the SQUIDs (see Methods).  $H_{\text{c}}=\sum_{l=1}^{N}g_l\left(\sigma_l^+ \hat{a}_l^\dagger + \sigma_l \hat{a}_l \right)+\sum_{l=1}^{N-1}J_{l}(t)\left(\hat{a}_l^\dagger \hat{a}_{l+1}^\dagger+\text{h.c.}\right)$ is the  counter-rotating term that can be neglected by using the rotating wave approximation (RWA).
	
The lowest three eigenstates of the JC Hamiltonian $h_l$ for the $l$th unit are $|0\text{g}\rangle_l $, $|$$\uparrow\rangle_l =\left(|0\text{e}\rangle_l + |1\text{g}\rangle_l \right)/\sqrt{2}$ and $|$$\downarrow \rangle_l =\left(|0\text{e}\rangle_l - |1\text{g}\rangle_l \right)/\sqrt{2}$, where $|n\text{g}\rangle_l$ and $|n\text{e}\rangle_l \; (n=0, 1, 2, \cdots)$ are the states containing $n$ photons while the transmon is at the ground and excited state, respectively. And their eigen-energies are $E_{l,0g}=0$ and $E_{l,\uparrow(\downarrow)}= \omega_l +(-) g_l$. Here, we exploit the two single-excitation  eigenstates $|\uparrow \rangle_l$ and $|\downarrow \rangle_l$ to simulate the effective electronic spin-up and spin-down state.  As each of the two states consists of half ``photon" and half ``atom", they are regarded as a whole and was termed as ``polariton''.

	\subsection{Polaritonic spin-orbit coupling}
We proceed to show that a spin-1/2 chain model with tunable Zeeman field and SOC can be simulated with the JC lattice by only adjusting the pulse shape of the coupling strengths $J_{l}(t)$ between neighbouring TLRs. Firstly, since each cell contains two pseudo-spin states, there are totally four inter-cell neighbouring hopping. In order to control each hopping separately, selective frequency addressing is employed (see Methods), i.e., we assign each of the four hopping with its unique hopping frequency. To achieve this, we adopt two sets of unit cells, A-type and B-type, which are different in the sense that they have different eigen-frequencies and coupling strengths of the JC model. Then we arrange them in an alternate way, as shown in Fig.~1A. Setting started with an A-type one,  when $l$ is odd (even), $\omega_l=\omega_{\text{A}}(\omega_{\text{B}})$ and $g_l=g_{\text{A}}(g_{\text{B}})$. Then,  based on the current experimental reaches \cite{SC3}, we set $\omega_{\text{A}}/2\pi= 6$ GHz, $\omega_{\text{B}}/2\pi= 5.65$ GHz, $g_{\text{A}}/2\pi=300$ MHz, and $g_{\text{B}}/2\pi=270$ MHz. With this, the energy intervals of the four hopping are $\{|E_{l,\alpha}-E_{l+1,\alpha'}|/2\pi\}=\{ 220, 320, 380, 920\}$ MHz with $\alpha, \alpha' \in \{\uparrow, \downarrow\}$. The differences between every two of them are no less than 20 times of the effective hopping strength $t_0/2\pi=3$ MHz, thus they can be selective addressed in frequency. Correspondingly, the $J_l(t)$ contains four tunes and can be written as
	\begin{equation}\label{coupling2}
	J_{l}(t)=\sum_{\alpha, \alpha'} 4t_{0,l\alpha\alpha'} \cos\left(\omega_{l\alpha\alpha'}^{\text{d}} t+  s_{l\alpha\alpha'} \varphi_{l\alpha\alpha'}\right),
	\end{equation}
	where
	\begin{equation}
		s_{l\alpha\alpha'}= \sgn(E_{l,\alpha}-E_{l+1,\alpha'})
		\label{sign1}
	\end{equation}
is the sign of the hopping phase, $4t_{0,l\alpha\alpha'},\; \omega_{l\alpha\alpha'}^{\text{d}}$ and $s_{l\alpha\alpha'} \varphi_{l\alpha\alpha'}$ are the amplitudes, frequencies and phases, corresponding to the hopping $|\alpha\rangle_l \rightarrow |\alpha'\rangle_{l+1}$, respectively. Note that due to the alternate arrangement, based on the definition above, the $J_l(t)$s take only two different forms -- when $l$ is odd (even), $J_l=J_{\text{AB}} (J_{\text{BA}})$ where the subscript ``AB" (``BA") refers to that the A-type (B-type) unit cell is on the left. Experimentally, this time-dependent coupling strength $J_l(t)$ can be realized by adding external magnetic fluxes with both dc and ac components threading through the SQUIDs (see Methods).  In this way, selective hopping can be induced, i.e., only when a frequency of the driving flux matches a particular hopping energy interval, that hopping can be triggered, otherwise it will not take into effect. Meanwhile, both the  strengths and phases  of the hopping can be controlled by the amplitudes and phases of the ac magnetic fluxes. However, only controlling these two is not enough for realizing the topological states, we still need to induce and adjust the spin splitting. Then we theoretically find out, and numerically prove that this spin splitting can be induced by just adding a detuning to the spin-flipped transition tunes. Concretely, while the spin-preserved transition frequencies are set as $\omega_{l\alpha\alpha}^{\text{d}}= |E_{l,\alpha}-E_{l+1,\alpha}|$,  we set the spin-flipped transition frequencies with a detuning $2m$  as  $\omega_{l\alpha\alpha'}^{\text{d}}= |E_{l,\alpha}- E_{l+1,\alpha'}| -2m$, where $\alpha\ne \alpha'$, as shown in Fig.~1B. Thus, in the rotating frame (explained later), there will be a spin splitting $m$ for each cell, as shown in Fig.~1C.

We now show how the time-dependent coupling strength in Eq. (\ref{coupling2}) can induce a designable spin transition process in a certain rotating frame. First, we map the Hamiltonian in Eq. (\ref{H_JC}) into the single excitation direct product subspace span$\{|0\text{g}, \cdots , 0\text{g}, \underset{l\text{th}}{\alpha}, 0\text{g}, \cdots, 0\text{g} \rangle \}_{l=1,\cdots,N;\, \alpha =\uparrow, \downarrow}$. Hereafter, when there is no ambiguity, we use $|\alpha \rangle_l$ to denote $|0\text{g}, \cdots, 0\text{g}, \underset{l\text{th}}{\alpha}, 0\text{g}, \cdots, 0\text{g}\rangle$, and $|G\rangle$ to denote $|0\text{g}, \cdots, 0\text{g} \rangle$. Then, we define a rotating frame by a unitary operator $U=\exp \{-\text{i}\left[\sum_{l} h_l-m(|\right.$$\uparrow$$\rangle_l\langle $$\uparrow$$ |- |$$\downarrow$$\rangle_l\langle $$\downarrow$$\left.  | )\right] t  \}$, which leads the Hamiltonian in Eq. (\ref{H_JC}) to $H'_{\text{JC}}=U^{\dag}H_{\text{JC}}U+\text{i}\dot{U}^{\dag}U$.  Neglecting the fast rotating terms (see Methods), one obtains
\begin{equation} \label{eq.simu}
H'_{\text{JC}}=\sum_{l}^{N} m \bm{S^z}_{l}
+\sum_{l=1}^{N-1} \sum_{\alpha, \alpha'} \left(t_{0,l\alpha\alpha'} \text{e}^{\text{i}\varphi_{l\alpha\alpha'}}
\hat{c}^\dagger_{l,\alpha} \hat{c}_{l+1,\alpha'} +\text{h.c.}\right),
\end{equation}
where $\bm{S^z}_{l}=|$$ \uparrow$$ \rangle_l\langle$$\uparrow$$|- |$$\downarrow$$\rangle_l\langle $$ \downarrow $$ |$, $\,t_{0,l\alpha\alpha'}$ is the effective coupling strength and $\hat{c}^\dagger_{l,\alpha}=|\alpha\rangle_l \, \langle G | $ is the creation operator for polariton with ``spin'' $\alpha$ in $l$th unit cell. As a result, the  Hamiltonian (\ref{eq.simu}) represents a general 1D spin-1/2 tight-binding lattice model with $m$, $t_{0,l\alpha\alpha'}$ and $\varphi_{l\alpha\alpha'}$  being the equivalent Zeeman energy,  hopping strength  and hopping phase, respectively. These three variables can all be experimentally tuned in wide ranges by  the frequencies, amplitudes and phases of the external ac magnetic fluxes. Notably, the effective on-site potential $m$ can be tuned as either positive or negative depending on the detuning direction.
	
Meanwhile, it is worth noticing that although we have introduced two kinds of unit cells (A-type and B-type) in the laboratory frame, by adjusting the coupling parameters $J_{\text{AB}}$ and $J_{\text{BA}}$,  the translation symmetry in the rotating frame is still preserved. In other words, the smallest repeating unit in the rotating frame contains only one unit cell as shown in Fig.~1C. By now, the adjustable spin-preserved tunneling  ($\alpha=\alpha'$)  and SOC terms ($\alpha \ne \alpha'$) can both be induced, hence our superconducting quantum circuit setup can naturally be used to simulate a tunable SOC topological polariton insulator.
	
In order to justify the individual frequency addressing of the inter-cell transitions, we  numerically simulate the dynamics of a system containing only two unit cells, with an A-type one on the left and a B-type one on the right. We test every hopping of the four transitions $\{|\alpha\rangle_{\text{A}} \leftrightarrow |\alpha'\rangle_{\text{B}}\}$ one by one, adding only one corresponding frequency $\omega_{1\alpha\alpha'}^{\text{d}}$ in $J_{\text{AB}}(t)$. As expected, when we pick out a resonant frequency $\omega_{1\alpha\alpha'}^{\text{d}}$ with $m=0$, there will be a Rabi oscillation between the two corresponding target states. One example of these Rabi oscillations was shown in Fig.~1D, which is of the least fidelity among the four.  Even in this worst case, and in the third Rabi cycle, the fidelity still reaches a high value of $0.9979$, which justify the RWA. In addition, our numerical simulation also shows that, in the present of the unmatched driving, all the initial non-target states remain almost unchanged, thus justify our individual frequency addressing method.

\subsection{Nodal-loop topological polaritons}
Our protocol provides a tunable platform using polaritons to study topological matters. Here, we take the nodal-loop semimetal as an application sample to demonstrate how to simulate a specific condensed matter model in our proposed setup. To fit our simulation setup, we  reform the Hamiltonian of the original 3D nodal-loop model  in Ref.~\cite{NodalLoop1}, without losing any topological properties. Firstly, we relabel the coordinates to set the hopping terms with SOC to be along the $x$ axis. Secondly, we consider the Fourier transformations along $y$ and $z$ directions with quasi-momenta $k_y$ and $k_z$ and treat them as system parameters.  Then, according to Eq.~(\ref{eq.simu}), we can simulate this 3D nodal-loop model in our 1D system with the other two dimensions being the parametric dimensions. In this direct simulation, to engineer the four transitions between different paloriton states, we need four different tunes in the inter-cell coupling strength $J_{l}^{\text{nod}}(t)$.

The above implementation can further be simplified as following. We first make a unitary transformation to the original Hamiltonian so that $H_\text{{nod}}=V^\dag H_\text{{ori}} V$,  where $V=\sum_{l=1}^{N} (-\text{i})^{l-1} \bm{I}_l$ with $\bm{I}_l=|$$\uparrow$$\rangle_l\langle $$\uparrow$$ |+ |$$\downarrow$$\rangle_l\langle $$\downarrow$$|$. The transformed 1D lattice Hamiltonian from Eq.~(\ref{eq.simu}), without losing any physical properties, is
\begin{align}	\label{H_nodal}
H_\text{{nod}}=&\sum_{l=1}^{N} 	m'(k_y, k_z)\bm{S^z}_{l}
+\sum_{l=1}^{N-1}\sum_{\alpha, \alpha'} \left(\text{i}t'_{0}\hat{c}^\dagger_{l,\uparrow} \hat{c}_{l+1,\uparrow}\right.\\
&\left.-\text{i}t'_{0}\hat{c}^\dagger_{l,\downarrow} \hat{c}_{l+1,\downarrow}+\text{i}t'_{0}\hat{c}^\dagger_{l,\uparrow} \hat{c}_{l+1,\downarrow}
-\text{i}t'_{0}\hat{c}^\dagger_{l,\downarrow} \hat{c}_{l+1,\uparrow} +\text{h.c.}\right),\notag
\end{align}
where $m'(k_y, k_z)=M+2d(\cos k_y+\cos k_z)$ with $M$ being the effective Zeeman energy and $d$ being the effective hopping energy along $y$ and $z$ directions.  Then, we set the parameters of the JC model to be $ \omega_{\text{A}}= \omega_{\text{B}}=2\pi \times 6$ GHz, $g_{\text{A}}/2\pi=200$ MHz and $g_{\text{B}}/2\pi=100$ MHz. Correspondingly, the coupling strengths can be set to contain only two tunes as
\begin{equation} \label{J_nod}
J_{l}^{\text{nod}}(t)=4t_0\cos\left[\omega_{1}^{\text{d}} t+\frac{(-1)^{l+1}\pi}{2}\right]+ 4t_0\cos\left(\omega_{2}^{\text{d}} t+ \frac{\pi}{2}\right),
\end{equation}
where $\omega_{1}^{\text{d}}/2\pi=100$ MHz and $\omega_{2}^{\text{d}}/2\pi=(300-2m)$ MHz. In this way, when transforming into the rotating frame of $U$ and applying the RWA, the Hamiltonian (\ref{H_JC}) of our JC-lattice system will takes the form of Eq.~(\ref{H_nodal}), which accomplishes the quantum simulation of  the topological nodal-loop semimetals (see Methods).

Experimentally, one can choose $t_0/2\pi=3$ MHz and set the detuning within the range of $m \in 2\pi \times [-20, 20] $ MHz, such that any value of the variable tunes $\omega_{l\alpha\alpha'}^d (\alpha \ne \alpha')$ still maintain a frequency difference no less than $20t_0$ away from other tunes, and thus the crosstalk caused by unwanted tunes will be negligible. Last, for testing the validity of our theoretical protocol, numerical simulations will be given in the Experimental detection section after a brief introduction of the characteristics of the topological nodal-loop polaritons.

\begin{figure*}[tb]
\centering
\includegraphics[width=11.5cm]{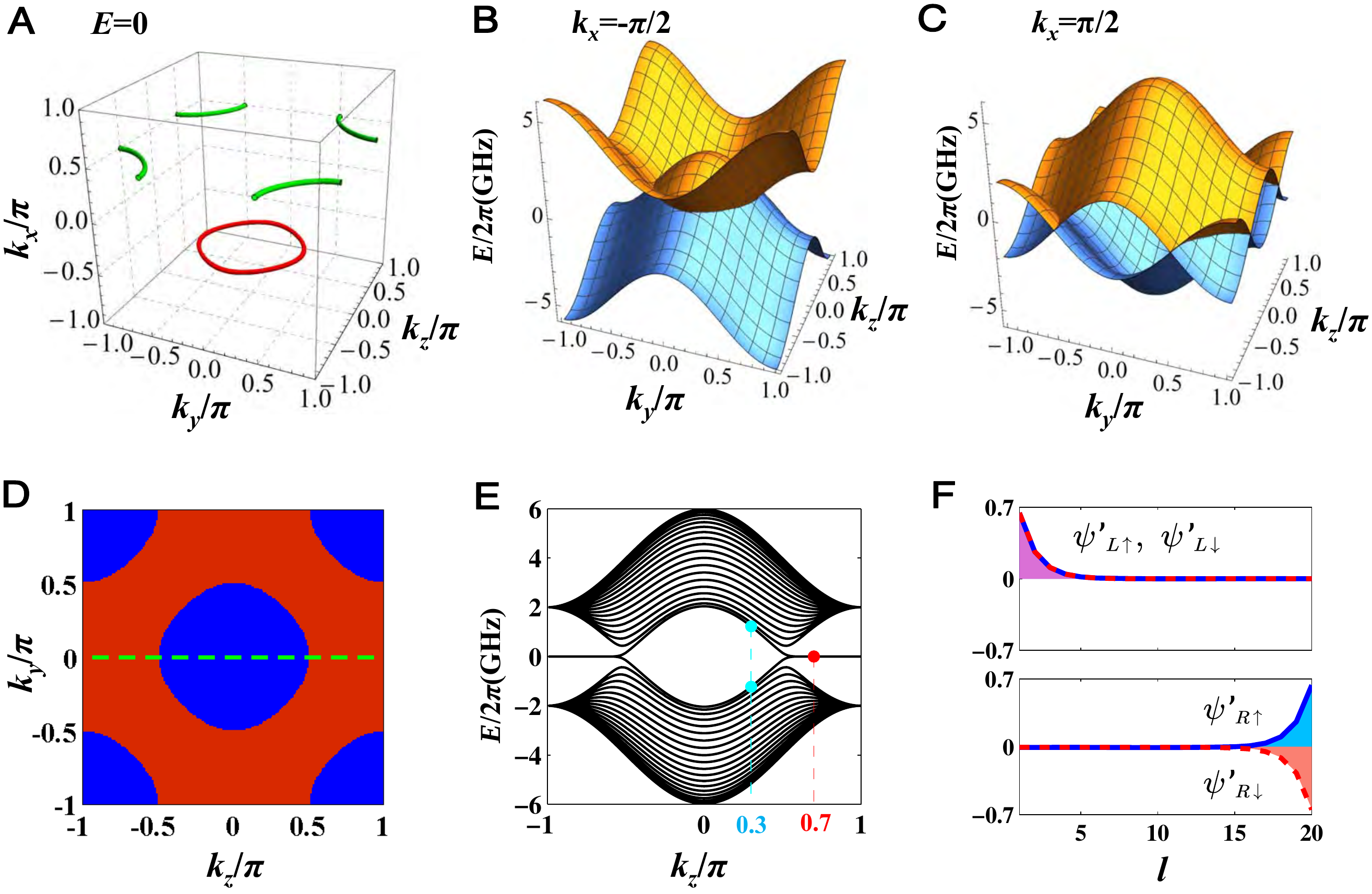}
\caption{\textbf{The band structure and topological characteristics of the simulated nodal-loop semimetal.} (A) The two loops (lines of $E=0$) where the two bands touch in the momentum space. There is one in the $k_x=-\pi/2$ plane (colored red) and one in the $k_x=\pi/2$ plane (colored green). The parameters for plotting are $M/t_0=0$ and $d/t_0=1$. Energy bands under the confines of (B) $k_x=-\pi/2$ and  (C) $k_x=\pi/2$. They touch each other along nodal loops. (D) The winding number altering over $k_y$ and $k_z$.  The red and blue regions are of winding number $\nu=1$ and $\nu=0$, respectively.  (E) Numerical calculation of the energy bands of a 1D 20-unit-cell lattice along $x$ direction in open boundary condition. The energies are altering over $k_z$ with confining $k_y$=0 (the dashed green line in D). The two light blue dots at $k_z=0.3\pi$ refers to two topological trivial states used for comparison in later discussion. The two red dots (overlapped) at $k_z=0.7\pi$ marks two in-gap zero-energy levels accompanied with two edge states whose wave functions are plotted in (F), where the blue and dashed red lines plot the probability $|\uparrow \rangle$ and $|\downarrow\rangle $ components  of the numerincally calculated wave-functions divided by a phase factor $\text{i}^{l-1}$, i.e., $\psi'_{L(R)\uparrow(\downarrow)}(l)=  \psi^{\text{num}}_{L(R) \uparrow(\downarrow)}(l)/\text{i}^{l-1}$.}
\label{nodal-loop}
\end{figure*}

\subsection{Characteristics of the topological polaritons}
	
To investigate the bulk characteristic of the topological nodal-loop polaritons, we first consider the periodic boundary condition to obtain the Hamiltonian in Eq. (\ref{H_nodal}) in the momentum space as
\begin{equation}
	H=b_y \bm{S^y}+ b_z\bm{S^z},
\end{equation}
where $b_y=-2t'_0 \cos k_x$, $b_z=2t'_0 \sin k_x+m'(k_y, k_z)$ and $\bm{S^{y}}=\text{i}( |\downarrow\rangle\langle\uparrow| -|\uparrow\rangle\langle\downarrow|)$. This system has two energy bands
	\begin{equation}
	E=\pm\sqrt{b_y^{2}+b_z^2},
	\label{energy}
	\end{equation}
which will touch when $E=0$. The touching points form closed lines, the so-called nodal-loops, in momentum space as shown in Fig.~2A. These two loops appear in the $k_x=\pi/2$ and/or $k_x=-\pi/2$ planes. By fixing $k_x=\pm\pi/2$, we plot these two energy bands in the $k_y$-$k_z$ space as shown in Fig.~2B and 2C, where one can see that the touching is right along the nodal-loops.
	
The topological index characterizing each nodal loop is a winding number  defined as
	\begin{equation}
	\begin{split}
	\nu(k_y, k_z) &=\frac{1}{2\pi }\int_{-\pi}^{\pi} dk_x \;  \mathbf{v}\times \partial _{k_{x}}\mathbf{v}\\
	&=\frac{1}{2}\big[\sgn\left( m'+2t'_0\right)
	-\sgn\left( m'-2t'_0\right) \big],
	\end{split}
	\label{gamma}
	\end{equation}
where $\mathbf{v}=(v_{y},v_{z})=(b_{y},b_{z})/\sqrt{b_{y}^{2}+b_{z}^{2}}$. This shows that the quantized winding number is either $1$ or $0$, corresponding to the cases whether a nodal loop is enclosing the straight line, which is along $k_x$ direction of the fixed $(k_x, k_y)$ point, or not \cite{NodalLoop1,NodalLoop2}. Hence, the nodal-loops divide the $k_y$-$k_z$ space into regions with different winding numbers, as shown in Fig.~2D. For all straight lines along $k_x$ inside the nodal loop, each of them can be regarded as being corresponding to a topological 1D gapped subsystem  with winding number $1$.

Two striking topological characters of nodal-loop semimetal are the zero-energy modes inside the energy gap and their corresponding edge states. We take a slice of $k_y=0$, indicated by the green line in Fig.~2D, as an example to plot the energy spectrum with various $k_z$ for a finite chain with $N=20$, as shown in Fig.~2E. The numerical result shows that there are two mid-gap degenerated zero-energy modes appear in the range of $\nu=1$ (the red area in Fig.~2D). The quantum states corresponding to the two mid-gap energies are edge states localized in the left and right end of the lattice, respectively. When $N$ is large enough for ignoring the finite-size effects, their wave-functions can be expressed as
\begin{subequations}\label{edge_state}
\begin{equation}
	\psi_{\text{L}} =\sum_{l=1}^{N} \text{i}^{l-1} A\text{e}^{-\lambda (l-1)} \left(|\uparrow\rangle_l+|\downarrow\rangle_l\right),
\end{equation}
\begin{equation}
	\psi_{\text{R}} =\sum_{l=1}^{N} \text{i}^{l-1} A\text{e}^{-\lambda (N-l)}\left(|\uparrow\rangle_l-|\downarrow\rangle_l\right),
\end{equation}
\end{subequations}
where $A=\sqrt{(1-q^2)/2(1-q^{2N})}$ and $\lambda=\ln(1/q)$ with $q=\tan(m'\pi/8t'_0)$. The phase factor $\text{i}^{l-1}$ stems from the unitary transformation $V$ to get the nodal-loop Hamiltonian in Eq. (\ref{H_nodal}) that simplified our simulation, from the original Hamiltonian in Eq. (\ref{eq.simu}). These analytical results of wave-functions are in very good agreement with the numerical results for $N=20$, as shown in Fig.~2F, thus justifies that we can use the JC lattice of experimentally capable size to simulate the topological features.

There are several phases in our simulated Hamiltonian where the phase transition is indicated by the emerging or vanishing of the nodal loops. Inferring from Eq.~(\ref{energy}), the critical conditions are obtained as $k_x=-\pi/2$ and $k_x=\pi/2$, which are corresponding to one nodal loop in each of the two regions of $-2t'_0-4d<M<-2t'_0+4d$ and $2t'_0-4d<M<2t'_0+4d$ with $d>0$ in the $M$-$d$  plane. Therefore, in the area where the two regions overlap, there are two nodal loops. But in the area outside these two regions, there is no nodal loop so that the whole Brillouin zone will be in a purely trivial or nontrivial phase. Consequently, there are totally five different phases of different winding number configurations $\nu(k_y, k_z)$ in the $M$-$d$ plane, as shown in Fig.~3. According to the chosen parameters, $t_0$ and $m$, the area $\{M,d |-6t'_0+4d<M<6t'_0-4d \}$ will include all the the five phases in the phase diagram of Fig.~3, i.e., all the possible phases can all be simulated in our proposed system.

\begin{figure}[tb]
\centering
\includegraphics[width=7.8cm]{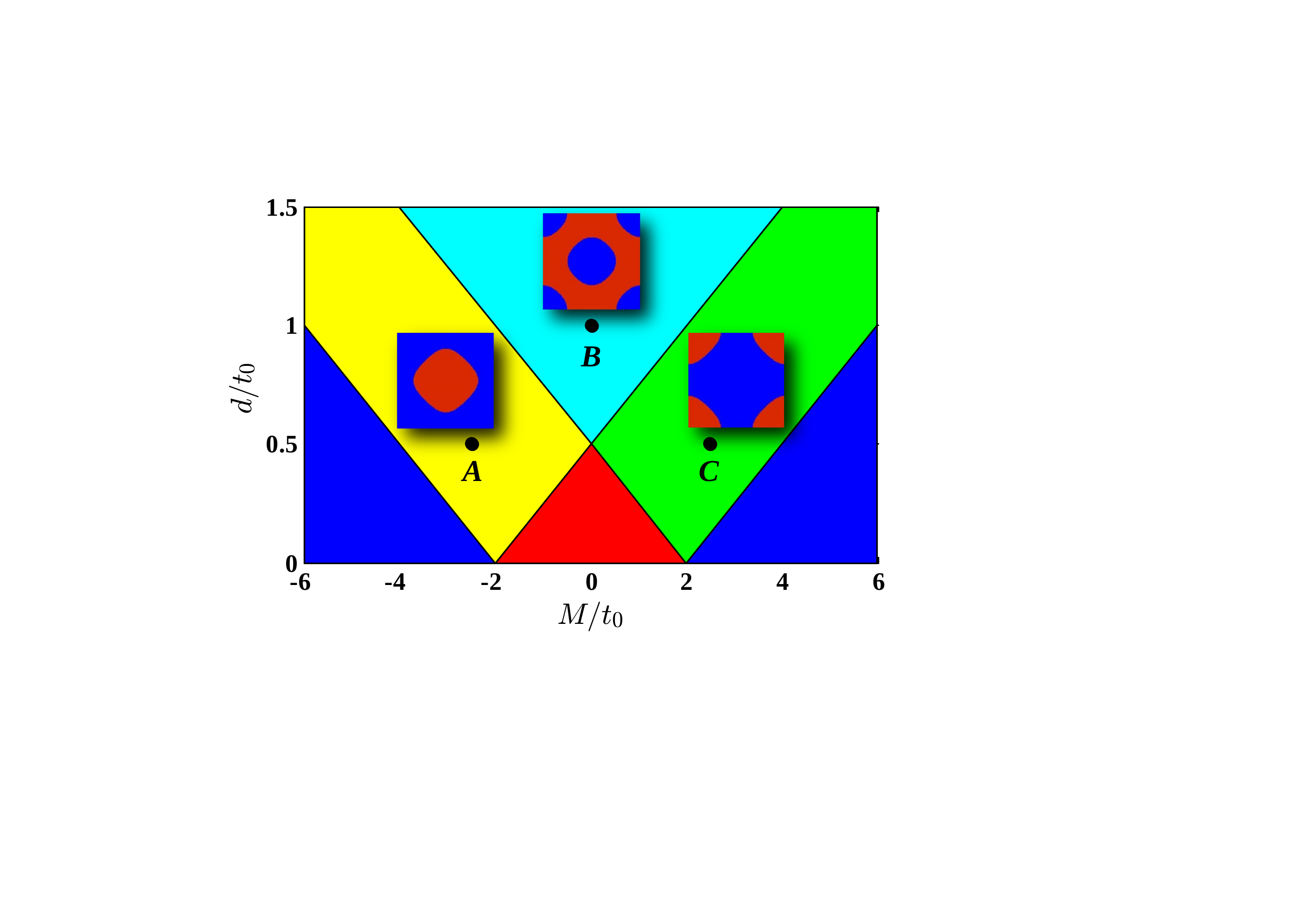}
\caption{\textbf{Phase diagram of nodal-loop semimetal bands with the corresponding winding number configurations.} Each color denotes a different phase. The dark blue region is of a trivial gapped phase without nodal loops in momentum space; the yellow and green region is of only one nodal loop in the $k_y$-$k_z$ plane; the light blue region is of two nodal loops in the $k_y$-$k_z$ plane; the red region is of a non-trivial gapped phase without nodal loops but with winding number $\nu=1$ in the whole $k_y$-$k_z$ plane. The three insets are the configuration of winding number in the $k_y$-$k_z$ plane (the red regions are of $\nu=1$, while blue regions are of $\nu=0$), corresponding to points A (-2.5, 0.5), B (0, 1) and C (2.5, 0.5), respectively.}
\label{phase}
\end{figure}

\subsection{Experimental detection methods}

\emph{Polaritonic topological edge state detection.} According to Eq.~(\ref{edge_state}), or as shown in Fig.~2F, the polariton in the left or right edge state is maximally distributed in the leftmost or rightmost JC lattice site. Their internal spins are in the superposition states  $\left(|\uparrow\rangle_l+|\downarrow\rangle_l \right)/\sqrt{2}$ and $\left(|\uparrow\rangle_l-|\downarrow\rangle_l \right)/\sqrt{2}$, respectively. Therefore one can find that, in the beginning, the left and right polaritonic edge states only have qubit excitation and photon components, respectively. Taking the detection of left edge state as an example, initially, the polariton in the leftmost JC lattice site is prepared into $|0\text{e}\rangle_1$, i.e., the leftmost qubit (resonator) has been prepared in the excited (vacuum) state. The qubits and resonators in the other sites are prepared into the ground and vacuum states which means the initial systematic state is $|\psi(t=0)\rangle=|0\text{e}\rangle_1|0\text{g}\rangle_2\cdots|0\text{g}\rangle_N$. After that, we let the above initial state evolve for a time about 0.5 $\mu$s. If the JC lattice is in the topological nontrivial phase supporting the left edge states, the final density distribution of the polaritons will maximally populate the leftmost site. The reason is that the initial state $|\psi(t=0)\rangle$ has a large overlap with the left edge state. It will evolve mainly via the edge state wave packet and maximally localized in the leftmost site. While if the system is in the topological trivial phase and has no edge states, the initial state will be a superposition of different bulk sates. The final density distribution will not  have maximal distribution in the leftmost site. Similarly, one also can prepare the JC lattice into $|\psi(t=0)\rangle=|0\text{g}\rangle_1\cdots|0\text{g}\rangle_{N-1}|1\text{g}\rangle_N$ and detect \cite{Lei-measu-wanghh} the right polaritonic topological edge state based on observing its time evolution.

In Fig.~4A and 4B, we have numerically calculated the time evolution of the polaritonic density when the JC lattice is in the topological trivial and nontrivial nodal-loop semimetal phases, respectively. For the trivial case, the wave packet has a ballistic spread versus time, which is a typical feature of bulk Bloch state. It shows that there is no edge state localization and the system is in topological trivial phase. For the nontrivial case, the density of the polaritons will always maximally localize in the leftmost JC lattice site, which indicates the existence of left topological edge states demonstrating that the system is in topological nontrivial phase. The time evolution of the qubit excitation and the photon population for the topological nontrivial case are also numerically calculated in Fig.~4C and 4D, which shows that the localized qubit excitation and the photon in the leftmost site have a Rabi-like oscillation feature inherited from the JC model.

\begin{figure}[tb]
\centering
\includegraphics[width=\columnwidth]{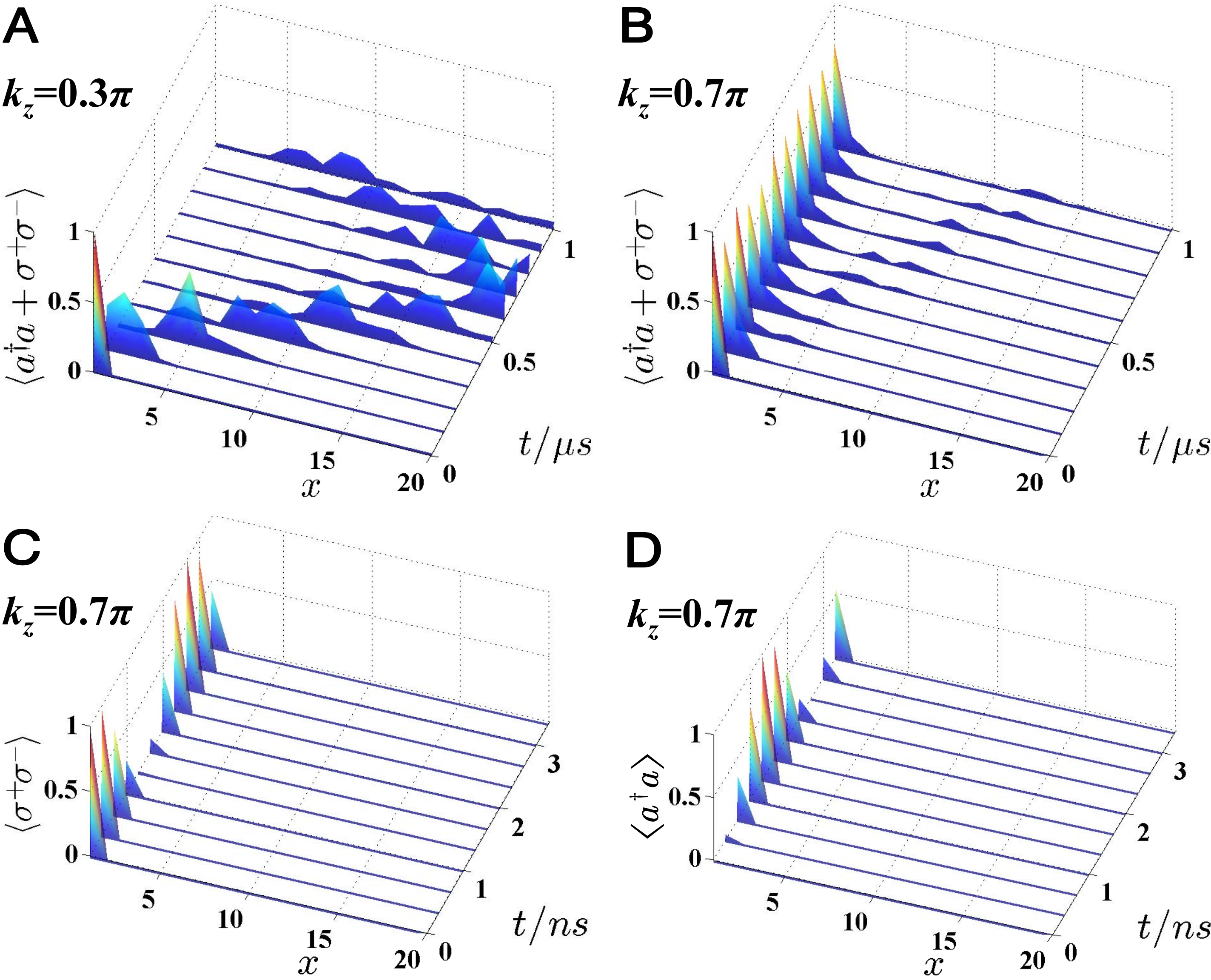}
\caption{\textbf{Dynamical detection of polaritonic topological edge states.} Time evolution of polaritonic density distribution $\langle \sigma^+\sigma^- + \hat{a}^\dagger \hat{a}\rangle$ when the JC lattice is in (A) topological trivial phase of $k_z=0.3\pi$ and (B) topological nontrivial phase of $k_z=0.7\pi$  (see Fig. 2E). Time evolution of (C) qubit excitation distribution $\langle \sigma_l^+\sigma_l^- \rangle$ and (D) photon distribution $\langle \hat{a}_l^\dagger \hat{a}_l\rangle$ for the topological nontrivial case. All the numerical simulations are based on the Hamiltonian in Eq. (\ref{H_JC}) without RWA. Other parameters are the same as that of in Fig.~2E.}
		\label{dynamic}
	\end{figure}	
	
\emph{Polaritonic topological invariant detection}. Another important hallmark for topological nontrivial nodal-loop polaritonic semimetal phase is the nontrivial polaritonic topological winding number. Here we show that such polaritonic topological invariant also can be dynamically detected. Our method is based on a previous work which shows that the topological winding number rotted in the momentum space can be detected through measuring the dynamical chiral center in the real space \cite{Mei}. The chiral operator for our topological polaritonic model is $\bm{S^x}_l=|$$\uparrow$$\rangle_l\langle$$\downarrow$$|+ |$$\downarrow$$\rangle_l\langle$$\uparrow$$|$. Then the chiral center operator for the JC lattice is defined as $\hat{P}_{\text{d}}=\sum_{l=1}^{N}l \bm{S^x}_l$. The polaritonic topological winding number can be related with the time-averaged dynamical chiral center associated with the single-polariton dynamics, i.e.,
	\begin{equation}
	\nu ={\lim_{T\rightarrow \infty }}\frac{2}{T}\int_{0}^{T}\text{d}t\, \langle \psi_{\text{c}}(t)| \hat{P}_{\text{d}} |\psi_{\text{c}}(t)\rangle,
	\label{pdt2}
	\end{equation}
where $T$ is the evolution time, $|\psi_{\text{c}}(t)\rangle =\exp(-\text{i}H_{\text{nod}}t)|\psi_{\text{c}}(0)\rangle$ is the time evolution of the initial single-polariton state $|\psi_{\text{c}}(0)\rangle=|0\text{g}\rangle_1\cdots |\uparrow\rangle_{\lceil N/2 \rceil} \cdots|0\text{g}\rangle_N$, where one of the middle JC lattice site has been put one polariton in, with its spin prepared in the state $|\uparrow\rangle$.

In Fig.~5A and 5B, we have numerically calculated the dynamical chiral center $\bar{P}_{\text{d}}(t)=\langle\psi_{\text{c}}(t)|\hat{P}_{\text{d}}|\psi_{\text{c}}(t)\rangle$ for topological trivial and nontrivial cases, respectively. According to Eq.~(\ref{pdt2}), one can find that the topological winding number is equal to twice the oscillation center of $\bar{P}_{\text{d}}(t)$. As shown in Fig.~5A, $\bar{P}_{\text{d}}(t)$ oscillates around the average value 0, which gives the polaritonic topological winding number $\nu=0$. In contrast, the result for the topological nontrivial case in Fig.~5B shows that $\bar{P}_{\text{d}}(t)$ oscillates around 0.5, which yields the polaritonic topological winding number $\nu=1$. Experimentally, the states of qubits and resonators can be measured with fidelity higher than $0.99$ in superconducting circuits. In our case, one only need to measure the qubit excitation and photon populations for getting their imbalance and deriving the chiral center, without requiring full quantum state tomography. In this way, the topological winding number can be easily and unambiguously detected based on monitoring single-polariton quantum dynamics in a JC lattice.
	
\begin{figure}[tb]
\centering
\includegraphics[width=\columnwidth]{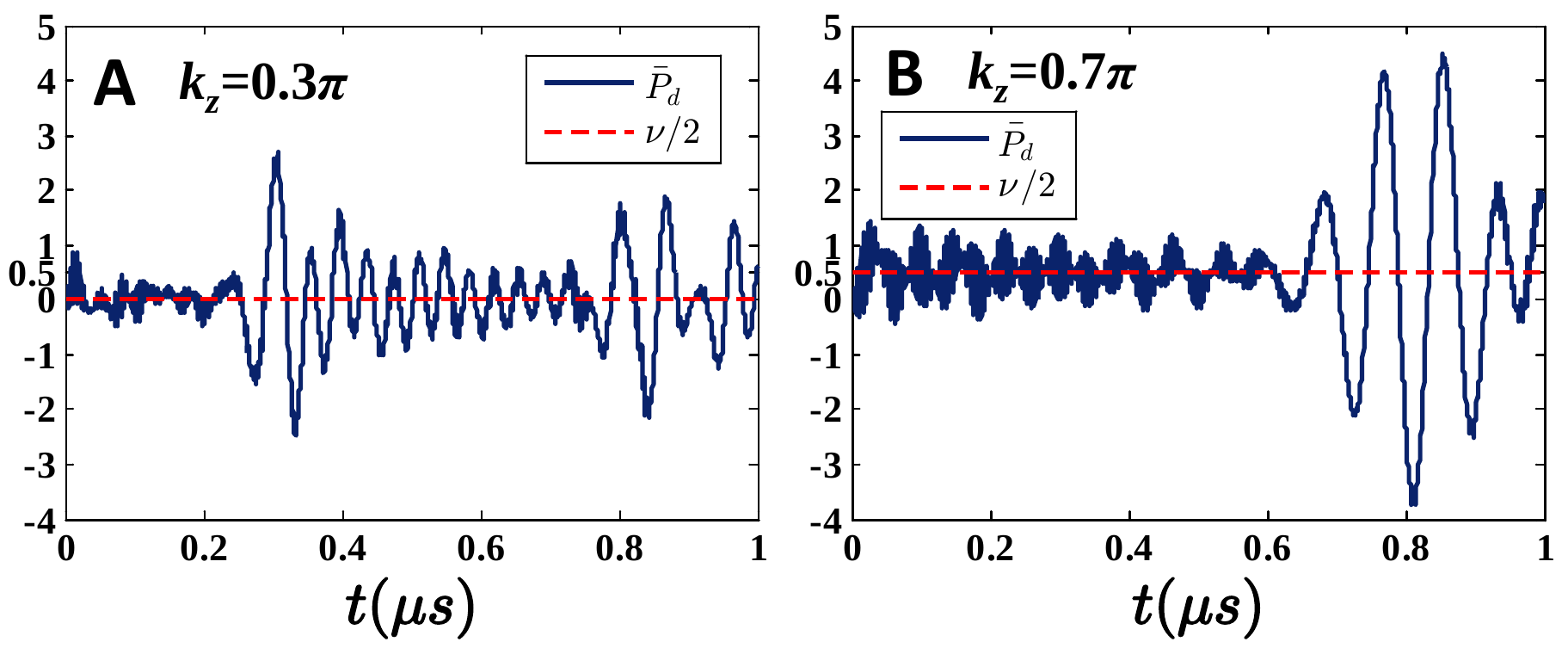}
\caption{\textbf{Dynamical detection of polaritonic topological invariants.} Time evolution of the chiral center $\bar{P}_{\text{d}}$ when the JC lattice is in (A) the topological trivial phase of $k_z=0.3\pi$ and (B) non-trivial phase of $k_z=0.7\pi$ (see Fig.~2E). The red dashed line denotes the oscillation center. All the numerical simulations are based on the Hamiltonian in Eq. (\ref{H_JC}) without RWA. Other parameters are chosen the same as that of in Fig.~4.}
\label{chiral}
\end{figure}

\section*{Discussion}

We further investigate how our construction and detection methods are influenced by the quantum decoherence effects. The Lindblad master equation is adopted to take three main decoherence factors including the losses of the photon and the decay and dephasing of the transmon into account. The Lindblad master equation of our system can be written as
\begin{equation}
    \dot {\rho }=-{\text{i}}[H_{\text{JC}},\rho ]+\sum_{l=1}^{N}\sum _{i=1}^{3}\gamma\left(\Gamma_{l,i}\,\rho \Gamma_{l,i}^{\dagger }-{\frac {1}{2}}\left\{\Gamma_{l,i}^{\dagger }\Gamma_{l,i},\rho \right\}\right),
\end{equation}
where $\rho$ is the density operator of the whole system, $\gamma$ is the decay rate or noise strength which are set to be the same here, $\Gamma_{l,1}=a_{l},\;\Gamma_{l,2}=\sigma^-_{l}$ and $\Gamma_{l,3}=\sigma^z_{l}$ are the photon-loss, transmon-loss  and the transmon-dephasing operators in the $l$th lattice, respectively. In Fig. 6A, we plot the edge-site population $P_1(t)=\text{tr}\left[\rho(t)\left(a_1^\dag a_1+\sigma_1^+\sigma_1^-\right)\right]$ after 0.5 $\mu$s and the oscillation center $\nu/2$ of the trivial and nontrivial cases for different decay rates. It shows that the edge state population and the chiral center smoothly decrease when the decay rate  increase. However, our detection method can tolerate the decay rate up to the order of $2\pi\times 100$ kHz, while the typical decay rates are only $2\pi\times 5$ kHz.

\begin{figure}[tb]
 \centering
\includegraphics[width=\columnwidth]{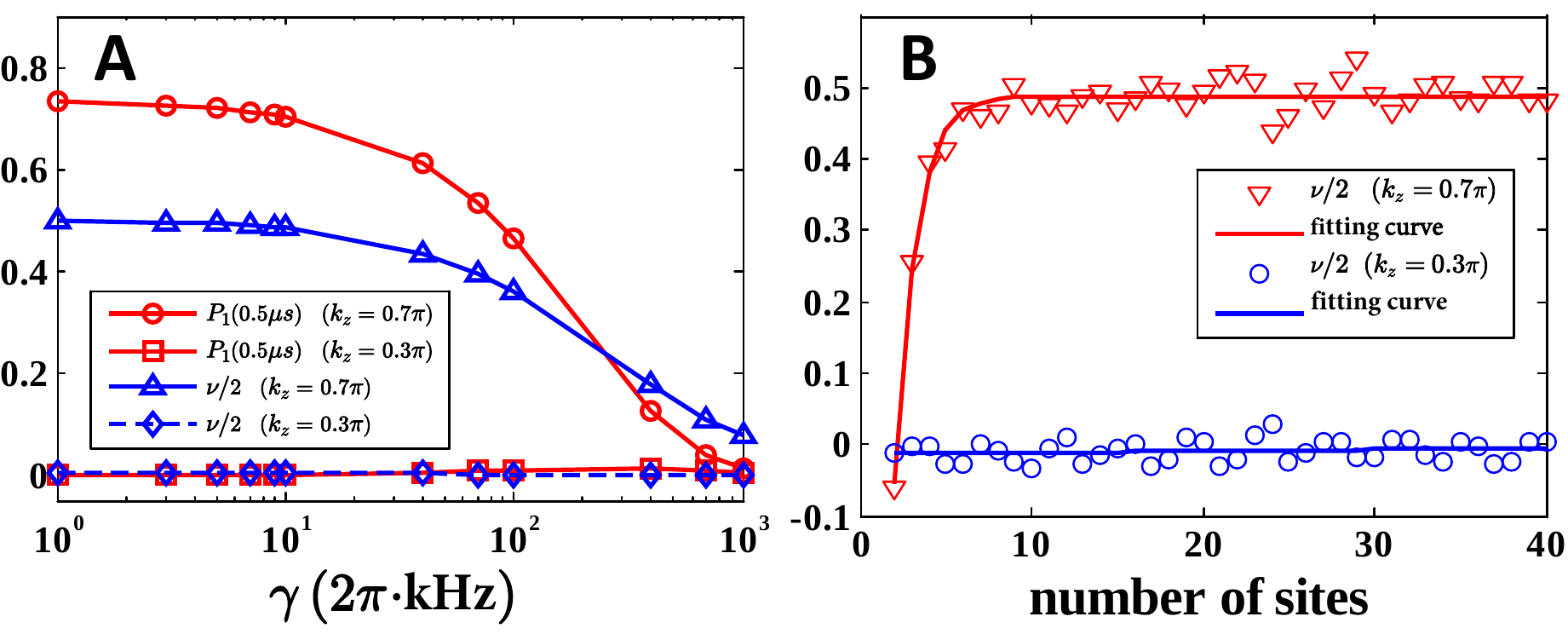}		
\caption{\textbf{The influences of the decoherence and the lattice site on the detection of the topological effects.} (A) The edge-site population $P_1(t)=\text{tr}\left[\rho(t)\left(a_1^\dag a_1+\sigma_1^+\sigma_1^-\right)\right]$ at 0.5 $\mu$s and the oscillation center $\nu/2$ of the topological trivial ($k_z=0.3\pi$) and nontrivial ($k_z=0.7\pi$) cases for different decay rate  $\gamma$. (B) The oscillation center of the topological trivial ($k_z=0.3\pi$) and nontrivial ($k_z=0.7\pi$) cases varying over the number of lattice sites. The decay rates are all set to be $2\pi\times 5$ kHz. All the numerical simulations are based on the Hamiltonian in Eq. (\ref{H_JC}) without RWA. Other parameters are chosen the same as that of in Fig.~4.}
\label{influences}
\end{figure}

We also investigate the minimum sites that are needed for the experimental detecting the oscillation-center. In the presence of the decay rates of $2\pi\times 5$ kHz, we plot the oscillation center for lattice with different number of sites in both the trivial and nontrivial cases, as shown in Fig. 6B. It shows that the four sites case corresponding to an oscillation center value about 0.40, is big enough for distinguishing the topologically trivial and nontrivial cases with current state-of-the-art technologies.

To conclude, we have introduced the concept of topological states into the JC lattice which is one of the most important building blocks in quantum optics and quantum information processing. We have studied the topological structure of light-matter interaction and shown that SOC physics and topological polaritons can be pursued in the JC lattice. Different from synthetic topological states in ultracold atomic, photonic and acoustic systems, topological polaritons are topological superposition states of photons and qubits. Tunable synthetic polaritonic SOC is induced by engineering the JC lattice couplings, which provides the basic ingredient for realizing spinful topological states of matter. We have also provided a method using single-particle quantum dynamics in real space to directly observe the polaritonic topological edge states and topological invariants.

Our work has a broad generalization and opens the door for exploring spinful topological states of matter using polaritons in JC lattice system. (i) In addition to mimicking spin-1/2, polariton states in multiple-excitation manifold have an extended spin degree of freedom and can also be used to mimic larger spin. It is challenging to realize SOC for the larger spin case in solid state materials and ultracold atoms. However, the method proposed in our work can be generalized to realize synthetic SOC for large-spin polaritons. This allows us to explore a large variety of topological states, including triple point topological states of matter \cite{TPS1}; (ii) Polariton states, the eigenstates of a JC model, can be referred as a synthetic dimension where each pair of states mimic a spin in a spin-lattice model and the coupling between polariton states provides the hopping in the synthetic dimension. With such synthetic dimension, high-dimensional topological states of matter can be explored in a low-dimensional JC lattice, including topological states beyond three dimensions \cite{4DTP1,4DTP2}; (iii) Polaritons have tunable strong nonlinear interaction, which allows us to study polaritonic fractional topological states of matter \cite{FTP1}; (iv) Besides superconducting circuit, JC lattices can also be realized in many quantum optical systems, including trapped ions \cite{Ion}, cavity quantum electrodynamics \cite{CQED}, nanoscopic lattice \cite{Nano}, optomechanical systems \cite{Opto} and so on.

\section{Methods}
	
\subsection{SQUID induced time-dependent photon hopping}
	
We here present how to induce the time-dependent photon hopping strength between two TLRs. As shown in Fig.~1A, neighbouring TLRs are connected by a common SQUID and an inductor $L$ in series then to the ground. The SQUID actually serves as a single Josephson junction (JJ) but with effective Josephson inductance tunable by the external flux. Concretely, by applying external magnetic flux  $\Phi_{\text{ext}}=\Phi_{\text{dc}}+\Phi_{\text{ac}}$  threading through the SQUID, when $\Phi_{\text{ac}} \ll \phi_0 $ with $\phi_{0}$ being the reduced flux quanta, the effective inductance of the SQUID reads \cite{Lei-induc3-prl}
	\begin{equation} \label{L_S}
	L_S(\Phi_{\text{ext}})=\frac{\phi_{0}}{2 I_{c}\cos \left(\Phi_{\text{ext}}/2\phi_{0} \right)},
	\end{equation}
where  $I_{\text{c}}$ is the shared critical current of the two JJ in each SQUID. On the other hand, comparing to the inductance of the TLRs, both the SQUID and the inductor $L$ have far smaller inductances, thus there is a voltage node but a current peak at both ends of each TLR. For these boundary conditions, after the conventional quantization of the TLRs \cite{Lei-induc3-prl,Nori-rew-Simu2-JC}, the flux density and the charge density wave-function of the lowest-energy mode in $l$th TLR can be expressed as
	\begin{subequations}
	\begin{equation}
	\hat{\phi}_l (x_l,t)=\frac{\sqrt{ \omega_l L_l} }{d_l} \cos\left(\frac{\pi}{d_l} x_l\right) \left[\hat{a}_l^\dagger (t)+ \hat{a}_l(t)\right],
	\end{equation}
	\begin{equation}
	\hat{q}_l(x_l,t)=\text{i}\frac{\sqrt{ \omega_l L_l} }{d_l} \sin\left(\frac{\pi}{d_l} x_l\right) \left[\hat{a}_l^\dagger(t)-\hat{a}_l(t) \right],
	\end{equation}
	\end{subequations}
where $\omega_l=\pi/\sqrt{L_lC_l}$ is the frequency of the photon with $L_l$ and $C_l$ being the inductance and capacitance, respectively. $d_l$ is the length and $x_l$ is the coordinate of the $l$th TLR. Meanwhile, because of the relatively low-impedances of the SQUID and the inductor, the currents from the ends of every TLR will flow directly through them to the ground,  without crossing to their neighbouring TLRs. Hence, the interaction Hamiltonian between the $l$th and $(l+1)$th TLR is just the summation energy of the SQUID and the inductor $L$
	\begin{eqnarray} \label{H_S}
	H_{\text{int}}^l&=&\frac{1}{2}(L_S+L)(I_l^{\text{ri}}+I_{l+1}^{\text{le}})^2 \notag\\
	&=& \sum_{j=l}^{l+1} \frac{\omega_j}{2L_j} \left( L_S+L \right) \left(\hat{a}_j^\dagger+\hat{a}_j\right)^2\\
&&	- \sqrt{\frac{ \omega_l \omega_{l+1}}{L_l L_{l+1}}} \left( L_S+L \right) \left(\hat{a}_l^\dagger+\hat{a}_l\right)\left(\hat{a}_{l+1}^\dagger+\hat{a}_{l+1}\right),\notag
	\end{eqnarray}
	where $I_l^{\text{ri(le)}}=\hat{\phi}_l (x_l,t) d_l /L_l|_{x_l=d_l(0)}$ is the current of right(left)-end of the $l$th TLR.
	Moreover, If we set
\begin{subequations}
\begin{equation}
	\Phi_{\text{ac}}=2\phi_{0}\arccos\frac{-1}{1+\sum_{j}^{n'}  \Omega_j \left[\cos\left(\omega_j^d t+\varphi_j\right)+1\right]} ,
	\end{equation}
	\begin{equation}
	t_{0,j}=\frac{ \phi_{0} \Omega_j}{8 I_{\text{c}} } \sqrt{ \frac{\omega_l\omega_{l+1}} {L_lL_{l+1}} },
	\end{equation}
	\begin{equation}
	L =\frac{\phi_{0}(\sum_{j}^{n'} \Omega_j+1)}{2 I_{\text{c}}},
	\end{equation}
	\end{subequations}
	where $n'$ is the number of the tunes in $\Phi_{\text{ac}}$, $\Phi_{\text{dc}}=4\pi n''\phi_0$ where $n''$ is an arbitrary positive integer, and choose resonant or detuned frequencies of $\omega_j$,  after the RWA, we obtain
	\begin{align} \label{HS3}
	H_{\text{int}}^l= 4\sum_{j}^{n'} t_{0,j}  \cos\left(\omega_j^d t+\varphi_j\right) \left(\hat{a}_l^\dagger\hat{a}_{l+1}+\text{h.c.}\right),
	\end{align}
	which right meets the form of Eqs. (\ref{H_JC}) and (\ref{coupling2}). This equation can be interpreted as describing the photons hopping between neighbouring unit cells, which means that the SQUID-L combination can actually serve as a counterpart of the semitransparent mirror in the optical cavity system.

	\subsection{Frequency addressing control}
We now show how the selective control of individual hopping in the JC lattice can be achieved by adjusting $J_l(t)$ in Eq.~(\ref{H_JC}) via the ac flux. We first map the Hamiltonian in Eq.~(\ref{H_JC}) into the single excitation subspace span$\{|\alpha\rangle_l\}_{\l=1, 2, \cdots, N; \; \alpha =\uparrow, \downarrow}$ and get
	\begin{equation}
	H_{\text{JC}}=\sum_{l,\alpha} E_{l,\alpha}\hat{c}_{l\alpha}^\dag \hat{c}_{l\alpha} +\frac{1}{2} \sum_{l=1}^{N-1} \sum_{\alpha, \alpha'} J_l(t) \hat{c}_{l,\alpha}^\dag \hat{c}_{l+1,\alpha'} + \text{h.c.},
	\end{equation}
where $\hat{c}_{l,\alpha}^\dag=|\alpha\rangle_l \langle G |$. Then in every $J_l(t)$, we add four tunes, corresponding to the four inter-cell hopping, cf. Eq.~(\ref{HS3}), each of which contains its independent tunable amplitude, frequency and phase as
	\begin{equation}
	J_l(t)=\sum_{\alpha,\alpha'} 4t_{0,l\alpha\alpha'} \cos\left(\omega_{l\alpha\alpha'}^d t+  s_{l\alpha\alpha'}\varphi_{l\alpha\alpha'}\right),
	\end{equation}
where $s_{l\alpha\alpha'}$ is defined in Eq.~(\ref{sign1}) being the sign of each phase and the frequencies are set as
	\begin{equation}
	\omega_{l\alpha\alpha'}^d=
	\begin{cases}
	|E_{l,\alpha}-E_{l+1,\alpha'}| \qquad \qquad(\alpha=\alpha'),\\
	|E_{l,\alpha}-E_{l+1,\alpha'}|-2m \quad \;\,(\alpha \ne \alpha'),
	\end{cases}
	\end{equation}
where $2m$ is a detuning. By now, the form of $J_l(t)$ is determined leaving $m$, $t_{0,l\alpha\alpha'}$ and $\varphi_{l\alpha\alpha'}$ to be chosen arbitrarily depending on the topological insulator model one simulates. Eventually, the target topological insulator model is in the rotating frame transformed by
\begin{equation}
U=\text{e}^{-\text{i}\left[\sum_{l} h_l-m(|\uparrow\rangle_l\langle \uparrow|- |\downarrow\rangle_l\langle \downarrow | )\right] t},
	\end{equation}
After the picture transformation $H'_{\text{JC}}=U^{\dag}H_{\text{JC}}U+\text{i}\dot{U}^{\dag}U$, one gets
	\begin{small}
\begin{align}
&H'_{\text{JC}}=\sum_{l=1}^{N} m \bm{S^z}_l +\sum_{l=1}^{N-1} \sum_{\alpha,\alpha'} t_{0,l\alpha\alpha'}\notag \\
&\times \left[\text{e}^{\text{i}(\omega_{l\alpha\alpha'}\,t+s_{l\alpha\alpha'}\varphi_{l\alpha\alpha'})} \right. + \left.
\text{e}^{-\text{i}(\omega_{l\alpha\alpha'}\,t+s_{l\alpha\alpha'}\varphi_{l\alpha\alpha'})} \right] \notag \\
&\times \left[\text{e}^{\text{i}(E_{l+1,\uparrow}-E_{l,\uparrow})t}\hat{c}_{l,\uparrow} \hat{c}^\dag_{l+1,\uparrow}+  % &
\text{e}^{\text{i}(E_{l+1,\downarrow}-E_{l,\downarrow})t}\hat{c}_{l,\downarrow}  \hat{c}^\dag_{l+1,\downarrow}\right.\notag \\
 & \quad + \left. \text{e}^{\text{i}(E_{l+1,\uparrow}-E_{l,\downarrow}-2m)t}\hat{c}_{l,\downarrow} \hat{c}^\dag_{l+1,\uparrow}\right. \notag \\
&\quad \left. + \text{e}^{\text{i}(E_{l+1,\downarrow}-E_{l,\uparrow}+2m)t}\hat{c}_{l,\uparrow} \hat{c}^\dag_{l+1,\downarrow}  \right]+\text{h.c.},
\end{align}\end{small}
	where $\bm{S^z}_{l}=|$$ \uparrow$$ \rangle_l\langle$$\uparrow$$|- |$$\downarrow$$\rangle_l\langle $$ \downarrow $$ |$.  After doing the multiplication in this equation, the four resonant terms and their Hermitian conjugates will be absent of time $t$ (frequency addressing).  Meanwhile, if the conditions $\{ t_{0,l\alpha\alpha'}\}_{\alpha,\alpha'=\uparrow,\downarrow}\ll  \{ \omega_{l\alpha\alpha'},\; \omega_{l\alpha\alpha'}- \omega_{l\beta\beta'} \}_{\alpha,\alpha',\beta,\beta' =\uparrow,\downarrow; \; \omega_{l\alpha\alpha'}\ne \omega_{l\beta\beta'} }$ are satisfied, all the other terms are fast rotating term that can be dropped within RWA. As a result, we obtain the tight-binding model with tunable SOC in Eq.~(\ref{eq.simu}).

    \subsection{Simplified method of implementation}
We here explain why the four hopping terms in the nodal-loop semimetal modal can be induced by the coupling strength in Eq.~(\ref{J_nod}) using only two tunes. Firstly, when we set parameters of the JC model to be $ \omega_{\text{A}}= \omega_{\text{B}}=2\pi \times 6$ GHz, $g_{\text{A}}/2\pi=200$ MHz and $g_{\text{B}}/2\pi=100$ MHz, the energy intervals of the four hopping overlap and reduce into two set of intervals, i.e., a spin-conserved hopping interval
    \begin{equation}
    |E_{l,\uparrow}-E_{l+1,\uparrow}|=|E_{l,\downarrow}-E_{l+1,\downarrow}|=2\pi \times 100 \text{MHz},
    \end{equation}
and a spin-flipped hopping interval
    \begin{equation}
    |E_{l,\uparrow}-E_{l+1,\downarrow}|= |E_{l,\downarrow}-E_{l+1,\uparrow}|= 2\pi \times 300  \text{MHz}.
    \end{equation}
Therefore, in this way, one tune in $J_{l}^{\text{nod}}(t)$ can induce two hopping while the two spin-conserved hopping still remain being controlled separated from the two spin-flipped hopping, thus they can be of different detuning.

Secondly, note that within this setting, according to the definition in Eq.~(\ref{sign1}), we have $s_{l\uparrow\uparrow}= -s_{l\downarrow\downarrow}$ and $s_{l\uparrow\downarrow}= -s_{l\downarrow\uparrow}$, thus both the two spin-conserved hopping terms and spin-flipped hopping terms can be induced by a same tune, but with opposite signs, i.e., $\text{i}$ and $-\text{i}$. This is possible because the RWA selects different terms. These opposite signs are ideal for realizing the wanted SOC in our protocol. Anyway, this is only possible when we do the unitary transformation $V$ to transform the hopping phase from the original $1$ and $-1$ into the pure imaginary numbers $\text{i}$ and $-\text{i}$. For the former case, one still has to use four tunes and induce them separately. Therefore, this transformation simplifies our simulation.

\section{ACKNOWLEDGEMENTS}
This work is supported by the NSFC (Grants No.~11874156,  No.~11604103 and No.~11604392),  the Key R\&D Program of Guangdong province (Grant No.~2018B0303326001), the National Key R\&D Program of China (Grant No.~2016YFA0301803 and No.~2017YFA0304203), the NSF of Guangdong Province (Grant No.~2016A030313436), the Startup Foundation of South China Normal University, the PCSIRT (Grant No.~IRT\_17R70), the 1331KSC, and the 111 Project (Grant No.~D18001).


\begin{thebibliography}{usrt}

\bibitem{JCModel} Cummings, F. W. \& Jaynes, E. T. Comparison of quantum and semiclassical radiation theories with application to the beam maser, \textit{Proceedings of the IEEE} \textbf{51} 89-109 (1963). %
		

\bibitem{Ion}  Leibfried, D., Blatt, R., Monroe, C. \& Wineland, D. Quantum dynamics of single trapped ions, \textit{Rev. Mod. Phys.} \textbf{75}, 281-324 (2003). %
		

\bibitem{CQED} Raimond, J. M., Brune, M. \& Haroche, S. Manipulating quantum entanglement with atoms and photons in a cavity, \textit{Rev. Mod. Phys.} \textbf{73}, 565-582 (2001). %
		

\bibitem{SC3} Devoret, M. H. \& Schoelkopf, R. J. Superconducting Circuits for Quantum Information: An Outlook, \textit{Science} \textbf{339}, 1169-1174 (2013). %
		

\bibitem{SC5} Xiang, Z. L., Ashhab, S., You, J. Q. \& Nori, F. Hybrid quantum circuits: Superconducting circuits interacting with other quantum systems, \textit{Rev. Mod. Phys.} \textbf{85}, 623-653 (2013). %
		

\bibitem{Nano} Chang, D. E., Douglas, J. S., Gonz\'{a}lez-Tudela, A., Hung, C.-L. \& Kimble, H. J. Quantum matter built from nanoscopic lattices of atoms and photons, \textit{Rev. Mod. Phys.} \textbf{90}, 031002 (2018) %
		

\bibitem{Opto} Aspelmeyer, M., Kippenberg, T. J. \& Marquardt, F. Cavity optomechanics, \textit{Rev. Mod. Phys.} \textbf{86}, 1391-1452 (2014). %
		

\bibitem{Hartmann} Hartmann, M. J., Brand\~{a}o, F. G. S. L. \& Plenio, M. B. Quantum many-body phenomena in coupled cavity arrays, \textit{Laser \& Photonics Rev.} \textbf{2}, 527-556 (2008). %
		

\bibitem{Fazio} Tomadin, A. \& Fazio, R.  Many-body phenomena in QED-cavity arrays,  \textit{J. Opt. Soc. Am. B} \textbf{27}, A130-A136 (2010).
		
	
\bibitem{Nori} Buluta, I. \& Nori, F. Quantum simulators,  \textit{Science} \textbf{326}, 108-111 (2009). %
		

\bibitem{Koch} Houck, A. A., T\"{u}reci, H.E. \& Koch, J. On-chip quantum simulation with superconducting circuits, \textit{Nat. Phys.} \textbf{8}, 292-299 (2012). %
		

\bibitem{Hartmann2006} Hartmann, M. J., Brand\~{a}o, F. G. S. L. \& Plenio, M. B. Strongly interacting polaritons in coupled arrays of cavities, \textit{Nat. Phys.} \textbf{2}, 849-855  (2006). %
		

\bibitem{Greentree} Greentree, A. D., Tahan, C., Cole,  J. H. \& Hollenberg, L. C. L. Quantum phase transitions of light, \textit{Nat. Phys.} \textbf{2}, 856-861 (2006). %
		
	
\bibitem{Angelakis} Angelakis, D. G., Santos, M. F. \& Bose, S. Photon-blockade-induced Mott transitions and XY spin models in coupled cavity arrays, \textit{Phys. Rev. A} \textbf{76}, 031805(R) (2007). %


\bibitem{add-njp} Nunnenkamp, A., Koch, J. \& Girvin, S. M. Synthetic gauge fields and homodyne transmission in Jaynes¨CCummings lattices, \textit{New J. Phys.} \textbf{13}, 095008 (2011).


\bibitem{add-prl109}Schir\'{o}, M., Bordyuh, M., \"{O} ztop, B. \& T\"{u}reci, H. E. Phase Transition of Light in Cavity QED Lattices, \textit{Phys. Rev. Lett.} \textbf{109}, 053601 (2012).


\bibitem{TPPho1} Lu, L., Joannopoulos, J. D. \& Solja\v{c}i\'{c}, M. Topological photonics, \textit{Nat. Photonics} \textbf{8}, 821-829 (2014).


\bibitem{TPPolariton} Karzig, T., Bardyn, C. E., Lindner, N. H. \& Refael, G. Topological Polaritons, \textit{Phys. Rev. X} \textbf{5}, 031001 (2015).

		

\bibitem{TPOpto} Peano, V., Brendel, C., Schmidt, M. \& Marquardt, F. Topological phases of sound and light, \textit{Phys. Rev. X} \textbf{5}, 031011 (2015).

		
\bibitem{TPCA1} Goldman, N., Budich, J. C. \& Zoller, P. Topological quantum matter with ultracold gases in optical lattices, \textit{Nat. Phys.} \textbf{12}, 639-645 (2016).


\bibitem{TPPhonon} Huber, S. D. Topological mechanics, \textit{Nat. Phys.} \textbf{12}, 621-623 (2016).
		

\bibitem{TPCA2} Cooper, N. R., Dalibard, J. \& Spielman, I. B. Topological bands for ultracold atoms, Rev. Mod. Phys. {\bf 91}, 015005 (2019).. %
		

\bibitem{TPCA3} Zhang, L. \& Liu, X. J. Spin-orbit coupling and topological phases for ultracold atoms, \textit{arXiv:}1806.05628 (2018). %
		
		

\bibitem{TPPho2} Ozawa, T. \textit{et al.} Topological photonics, Rev. Mod. Phys. {\bf 91}, 015006 (2019).
%, H. M. Price, A. Amo, N. Goldman, M. Hafezi, L. Lu, M. Rechtsman, D. Schuster, J. Simon, O. Zilberberg, I. Carusotto,
		
		

\bibitem{Kane} Hasan, M. Z. \& Kane, C. L. Topological insulators, \textit{Rev. Mod. Phys.} \textbf{82}, 3045-3067 (2010). %
		

\bibitem{Zhang} Qi, X. L. \& Zhang, S. C. Topological insulators and superconductors, \textit{Rev. Mod. Phys.} \textbf{83}, 1057-1110 (2011). %
		

\bibitem{TPQC} Nayak, C., Simon, S. H., Stern, A., Freedman, M. \& Sarma, S. D. Non-Abelian anyons and topological quantum computation, \textit{Rev. Mod. Phys.} \textbf{80}, 1083-1159 (2008). %
		

\bibitem{SOC1} Dalibard, J., Gerbier, F., Juzeli\={u}nas, G., \& \"{O}hberg, P. Artificial gauge potentials for neutral atoms, \textit{Rev. Mod. Phys.} \textbf{83}, 1523-1543 (2011). %
		

\bibitem{SOC2} Galitski, V. \& Spielman, I. B. Spin-orbit coupling in quantum gases, \textit{Nature (London)} \textbf{494}, 49-54 (2013). %
		

\bibitem{SOC3} Goldman, N., Juzeli\={u}nas, G., \"{O}hberg, P. \& Spielman, I. B. Light-induced gauge fields for ultracold atoms, \textit{Rep. Prog. Phys.} \textbf{77}, 126401 (2014). %
		

\bibitem{TPExp2} Jotzu, G. \textit{et al.} Experimental realization of the topological Haldane model with ultracold fermions, \textit{Nature (London)} \textbf{515}, 237-240 (2014). %M. Messer, R. Desbuquois, M. Lebrat, T. Uehlinger, D. Greif, T. Esslinger,
		

\bibitem{TPExp3} Aidelsburger, M. \textit{et al.}  Measuring the Chern number of Hofstadter bands with ultracold bosonic atoms, \textit{Nat. Phys.} \textbf{11}, 162-166 (2015). %M. Lohse, C. Schweizer, M. Atala, J. T. Barreiro, S. Nascimb\`{e}ne, N. R. Cooper, I. Bloch, N. Goldman,
		


\bibitem{TPExp4} Wu, Z. \textit{et al.} Realization of two-dimensional spin-orbit coupling for Bose-Einstein condensates, \textit{Science} \textbf{354}, 83-88 (2016). %, L. Zhang, W. Sun, X. T. Xu, B. Z. Wang, S. Ji, Y. Deng, S. Chen, X. Liu, J.  Pan,
		

\bibitem{TPExp6} Fl\"{a}schner, N. \textit{et al.} Observation of dynamical vortices after quenches in a system with topology, \textit{Nat. Phys.} \textbf{14}, 265-268 (2018). %D. Vogel, M. Tarnowski, B. S. Rem, D.-S. L\"{u}hmann, M. Heyl, J. C. Budich, L. Mathey, K. Sengstock, C. Weitenberg,
		

\bibitem{NLS} Bzdu\v{s}ek, T., Wu, Q., R\"{u}egg, A., Sigrist, M. \& Soluyanov, A. A. Nodal-chain metals, \textit{Nature (London)} \textbf{538}, 75-78 (2016).
		
	
		
\bibitem{NodalLoop1} Burkov, A. A., Hook, M. D. \& Balents, L. Topological nodal semimetals, \textit{Phys. Rev. B} \textbf{84}, 235126 (2011).
		

\bibitem{NodalLoop2} Zhang, D.-W. \textit{et al.} Quantum simulation of exotic PT-invariant topological nodal loop bands with ultracold atoms in an optical lattice, \textit{Phys. Rev. A} {\bf 93}, 043617 (2016).	%,  Y.-X. Zhao, R.-B. Liu, Z.-Y. Xue, S.-L, Zhu, Z. D. Wang,
		
		
\bibitem{Nori-rew-Simu2-JC}Gu, X., Kockum, A. F., Miranowicz, A., Liu, Y.-X. \& Nori, F.  Microwave photonics with superconducting quantum circuits, \textit{Phys. Rep.} \textbf{718-719}, 1-102 (2017). %


\bibitem{Lei-measu-wanghh} LinPeng, X. Y. \textit{et al.} Joint quantum state tomography of an entangled qubit-resonator hybrid, \textit{New J. Phys.} \textbf{15}, 125027 (2013).% H. Z. Zhang, K. Xu, C. Y. Li, Y. P. Zhong, Z. L. Wang, H. Wang, Q. W. Xie,
		

\bibitem{Mei} Mei, F., Chen, G., Tian, L., Zhu, S. L. \& Jia, S.  Topology-dependent quantum dynamics and entanglement-dependent topological pumping in superconducting qubit chains, \textit{Phys. Rev. A} \textbf{98}, 032323 (2018).
		

\bibitem{TPS1} Bradlyn, B. \textit{et al.} Beyond Dirac Weyl fermions: Unconventional quasiparticles in conventional crystals, \textit{Science} \textbf{353}, aaf5037 (2016). %, J. Cano, Z. Wang, M. G. Vergniory, C. Felser, R. J. Cava, B. A. Bernevig,
		

\bibitem{4DTP1} Lohse, M., Schweizer, C., Price, H. M., Zilberberg, O. \& Bloch, I. Exploring 4D quantum Hall physics with a 2D topological charge pump, \textit{Nature (London)} \textbf{553}, 55-58 (2018). %
		

\bibitem{4DTP2} Zilberberg, O. \textit{et al.} Photonic topological boundary pumping as a probe of 4D quantum Hall physics, \textit{Nature (London)} \textbf{553}, 59-62 (2018). %, S. Huang, J. Guglielmon, M. Wang, K. P. Chen, Y. E. Kraus, M. C. Rechtsman,
		

\bibitem{FTP1} Hayward, A. L. C., Martin, A. M. \& Greentree, A. D. Fractional Quantum Hall Physics in Jaynes-Cummings-Hubbard Lattices, \textit{Phys. Rev. Lett.} \textbf{108}, 223602 (2012). %
		

\bibitem{Lei-induc3-prl} Felicetti, S. \textit{et al.} Dynamical Casimir Effect Entangles Artificial Atoms, \textit{Phys. Rev. Lett.} \textbf{113}, 093602 (2014).%M. Sanz, L. Lamata, G. Romero, G. Johansson, P. Delsing, E. Solano,


		
	\end{thebibliography}
\end{document}